\documentclass[conference]{IEEEtran}
\IEEEoverridecommandlockouts
\usepackage{cite}
\usepackage{amsmath,amssymb,amsfonts}
\usepackage{url}
\usepackage{booktabs}
\usepackage{algorithmic}
\usepackage{graphicx}
\usepackage{textcomp}
\usepackage{rotating}
\usepackage{xcolor}
\def\BibTeX{{\rm B\kern-.05em{\sc i\kern-.025em b}\kern-.08em
    T\kern-.1667em\lower.7ex\hbox{E}\kern-.125emX}}
\begin{document}

\title{Locational Marginal Pricing of Compute:
Co-optimizing Compute and Energy for Transparent Data
Center Flexibility

}


\author{\IEEEauthorblockN{Vladimir Pyltsov}

\IEEEauthorblockA{
\textit{Columbia University}\\
New York, NY \\
v.pyltsov@columbia.edu}
\and
\IEEEauthorblockN{Teng Zeng}
\IEEEauthorblockA{\textit{Independent Researcher} \\
San Francisco, CA \\
tengzeng@berkeley.edu}

}

\maketitle

\begin{abstract}
Demand from data centers and AI compute is growing rapidly and is increasing stress on the power grid. This load could be flexible through user adjustments and provide relief during energy scarcity events, yet today its flexibility is provided through utility call-ups or system-operator load shedding, and is invisible to the users whose workloads are being curtailed. In this work, we propose to quantify and couple the prices of compute and energy in a single optimization problem. The resulting compute price reflects the marginal cost of the binding constraints and decomposes into components indicating which grid, power, hardware, or service requirement causes a price surge. The economic interpretation updates the price with real-time conditions and gives users a signal for potential flexibility provision. The co-optimization ultimately connects compute and energy markets while providing a signal for relief of stress on both the grid and the compute resources.

\end{abstract}

\begin{IEEEkeywords}
Data Center, Compute Pricing, Flexibility 
\end{IEEEkeywords}

\section{Introduction}

Data center and AI compute demand has been growing at a rapid pace. The projections forecast this demand to make up to 12\% of the U.S. electricity demand by 2028 \cite{shehabi20242024} and increase two-fold globally by 2030 \cite{chen2025electricity}. Such substantial growth can create stress on the grid system, impacting grid infrastructure and cost allocation \cite{cheng2026ai} as well as electricity prices \cite{bogmans2026power}.

Recently, data centers have been argued to be a flexible load, relieving the system during stress periods. A variety of works provide proposals for temporal and spatial workload shifting \cite{chen2026defer, sukprasert2024limitations, radovanovic2022carbon}. Recent work demonstrates the effectiveness of power cap management through the utility event call-up \cite{colangelo2026ai} and coordination between the utility and data center sides \cite{fatima2026unlocking}.

Yet, the flexibility approaches proposed are limited. In the current configuration, users are not exposed to the operational conditions of the serving infrastructure, nor have an incentive to provide demand response. The majority of the current pricing contracts can be described with the following attributes: a) constant price throughout a day with very slow movement on an interday basis; b) constrained on several compute metrics (i.e., promised deliverable time and available capacity); c) usually billed on a \$/GPU-hr or \$/token basis. 

While the customers are shielded from the volatility of operating conditions and obtain service on promise, this prevents users from providing valuable flexibility. In this work, we propose the co-optimization approach between the grid and compute pricing. The marginal compute price, which arises as a dual of the binding infrastructure and serving requirement constraints, becomes a decomposable quantity directly linked to the operational conditions. This provides transparency for customers in terms of which of their compute requirements creates stress on the current system, giving leverage to relax or defer the workload.  



\subsection{Related Work}

Prior work on data center flexibility falls into three groups. Geographic load balancing routes requests across sites to lower the operator's cost under exogenous electricity prices \cite{rao2010minimizing,liu2011greening,liu2013data}. A second group places data centers inside the grid model and guides spatial and temporal shifting with LMPs and market clearing \cite{lindberg2021guide,lindberg2022using,lin2021evaluating,zhang2020flexibility,zhang2022remunerating,gu2025competitive}. These works price electricity and the flexibility to shift it, but not compute. A third group co-optimizes compute and power within the fleet, using DVFS, storage, replica commitment, or iteration with the system operator \cite{liu2026watts,du2026spatial,ding2026towards,liu2026peer}, yet users still pay a fixed service price. Compute pricing has emerged only recently. Billimoria and Byers \cite{billimoria2026bits} infer the value of compute from posted cloud prices, and Sanabria \cite{sanabria2026shadow} derives a shadow price of LLM queries on a single system. Closest to ours, Liu \cite{liu2026locational} decomposes a dual \$/token price over a token-flow network, but with exogenous electricity prices and fixed energy per token.

Table \ref{tab:related} summarizes the comparison. To our knowledge, no prior work forms the compute price as a dual of a joint clearing with an endogenous grid. Hence, our contributions are as follows: (i) we couple compute demand and data center power demand in a single clearing problem; (ii) we form the compute price transparently as the dual of the binding constraints, decomposed into components that reflect actual grid, power, hardware, and service conditions; and (iii) we extend co-optimization from electricity to joint compute and energy clearing, giving a benchmark toward which system and fleet operators could coordinate by exchanging prices and quantities.


\begin{table}[t]
\centering\small
\caption{Comparison with the related work. \checkmark: addressed; ($\checkmark$):
partially or as an extension; --: not addressed.}
\label{tab:related}
\setlength{\tabcolsep}{3pt}
\resizebox{\linewidth}{!}{%
\begin{tabular}{@{}lcccccc@{}}
\toprule
 & \rotatebox{70}{Compute price}
 & \rotatebox{70}{Grid endogenous (OPF)}
 & \rotatebox{70}{Network/latency}
 & \rotatebox{70}{Operating modes (batch size)}
 & \rotatebox{70}{SLO reserve (queueing)}
 & \rotatebox{70}{Elastic demand} \\
\midrule
Rao et al.~\cite{rao2010minimizing}; Liu et al.~\cite{liu2011greening,liu2013data}
  & -- & -- & \checkmark & -- & ($\checkmark$) & -- \\
Lindberg et al.~\cite{lindberg2021guide,lindberg2022using}
  & -- & \checkmark & -- & -- & -- & -- \\
Lin et al.~\cite{lin2021evaluating}
  & -- & \checkmark & -- & -- & -- & -- \\
Zhang et al.~\cite{zhang2020flexibility,zhang2022remunerating}
  & -- & \checkmark & -- & -- & -- & \checkmark\\
Gu and Qin~\cite{gu2025competitive}            
  & -- & \checkmark & -- & -- & -- & --  \\
Ding et al.~\cite{ding2026towards}              
  & -- & \checkmark & -- & -- & -- & -- \\
Liu et al.~\cite{liu2026peer}
  & ($\checkmark$) & \checkmark & -- & -- & \checkmark & --  \\
Liu, Shin, and Deka~\cite{liu2026watts}
  & -- & -- & -- & \checkmark & -- & ($\checkmark$)  \\
Du et al.~\cite{du2026spatial}
  & -- & -- & \checkmark & -- & ($\checkmark$) & --  \\
Billimoria and Byers~\cite{billimoria2026bits}
  & ($\checkmark$) & -- & -- & -- & -- & \checkmark  \\
Sanabria~\cite{sanabria2026shadow}
  & \checkmark & -- & -- & ($\checkmark$) & ($\checkmark$) & ($\checkmark$)  \\
Liu~\cite{liu2026locational}
  & \checkmark & -- & \checkmark & -- & -- & ($\checkmark$)  \\
\textbf{This work}
  & \checkmark & \checkmark & \checkmark & \checkmark & \checkmark & \checkmark\\
\bottomrule
\end{tabular}}
\end{table}

\section{Methodology}\label{sec:method}

We clear compute and electricity in a single co-optimization problem. The power side is a DC optimal power flow (DC-OPF), whose duals are the locational marginal prices (LMPs) of electricity, i.e., the energy price plus a congestion component. The compute side serves requests from demand regions $r\in\mathcal{R}$ at data-center sites $s\in\mathcal{S}$; its duals are the prices of compute. The two sides are coupled because every request served at site $s$ is a power withdrawal at the bus $n(s)$ of that site. The intervals are hourly at $\tau=3600$ seconds. 

\subsection{Requests as the unit of demand}

An LLM request consists of an input (prompt and context, processed in the prefill phase) and an output (generated in the decode phase). The two phases utilize the hardware differently: prefill is compute-bound and decode is memory-bound. However, a request has no value unless both are served. We therefore measure demand in requests and let each request consume both prefill and decode resources.


We create different request classes $k\in\mathcal{K}=\{\text{conversation},\text{agentic},\text{offline}\}$ that differ in their latency requirements. Each class has an average $n^{\rm in}_k$ input and $n^{\rm out}_k$ output tokens per request and a request rate $D_{krt}$ (requests/s) from region $r$. We assume elastic demand, represented by bid segments on a demand curve. We split $D_{krt}$ into $N_B$ bid segments $b$ with willingness to pay $p_{kb}$ per token. The prices are spaced geometrically between the lowest and highest posted prices, $p^{\rm lo}_k$ and $p^{\rm hi}_k$, for the same model. Assuming a constant price elasticity $\varepsilon$ (1 in our case), the share of demand still buying at price $p$ is $(p/p^{\rm lo}_k)^{-\varepsilon}$. The segment share $\alpha_{kb}$ is the drop in this share between consecutive prices, and the top segment keeps the remainder ($\sum_b\alpha_{kb}=1$). A request in segment $b$ is worth its token count times the segment price:
\begin{equation}
  v_{kb}=p_{kb}\big(n^{\rm in}_k+n^{\rm out}_k\big)\qquad
  \bar d_{krtb}=\alpha_{kb}\,D_{krt} \label{eq:bid}
\end{equation}
We also assume the bids to be identical for input and output tokens. 

\subsection{Resources used by one request}

Prefill is modeled by one throughput $\theta^{\rm pre}$ (input tokens/s per GPU) and one energy $e^{\rm pre}$ per input token. Decode is modeled by operating modes $m\in\mathcal{M}$, which depend on batch size $B_m$. This results in different output throughput $\theta_m$ (tokens/s per GPU), an energy per output token $e_m$, and a 95th-percentile inter-token latency (ITL) $\ell_m$. Larger batches give more tokens per GPU and less energy per token, but slower generation. Serving a class-$k$ request in mode $m$ takes $a_{km}$ GPU-seconds and $e_{km}$ energy:
\begin{align}
  a_{km} &= a^{\rm in}_k + a^{\rm out}_{km}
          = \frac{n^{\rm in}_k}{\theta^{\rm pre}} + \frac{n^{\rm out}_k}{\theta_m} \label{eq:a}\\
  e_{km} &= e^{\rm in}_k + e^{\rm out}_{km}
          = n^{\rm in}_k e^{\rm pre} + n^{\rm out}_k e_m \label{eq:e}
\end{align}

A mode may serve class $k$ from region $r$ at site $s$ only if it meets the ITL target $L_k$, the network round-trip budget $R_k$, and the GPU memory limit. The memory limit comes from the proxy model of the key-value (KV) cache. A replica with $M$ bytes of free memory and $\kappa$ bytes of cache per token holds at most $B^{\max}_k$ sequences:
\begin{equation}
  B^{\max}_k=\frac{M}{\kappa\,\big(n^{\rm in}_k+n^{\rm out}_k/2\big)}
\end{equation}
We assume that each running sequence holds its input and, on average, half of its output. The admissible set is
\begin{equation}
\begin{aligned}
  \mathcal{A}=\big\{(k,r,s,m):\ &\ell_m\le L_k,\ B_m\le B^{\max}_k,\\
  &\mathrm{RTT}_{rs}\le R_k\big\}
\end{aligned}\label{eq:adm}
\end{equation}


For each class, $\rho_k$ is the highest utilization at which $\geq95\%$ of requests wait less than $W_k$.



\subsection{Clearing problem}

The decision variables are: 1) the served request rate $z_{krsmt}\ge0$ for each $(k,r,s,m)\in\mathcal{A}$; 2) the accepted demand $d_{krtb}\ge0$; 3) the powered-on GPUs $y_{st}\ge0$; 4) the site power $P_{st}$; 5) the generator outputs $g_{it}$; 6) the load shedding $q_{nt}\ge0$ of other customers at bus $n$.


The net injection at bus $n$ is:
\begin{equation}
  p_{nt}=\sum_{i\in\mathcal{G}_n}g_{it}-D^{\rm grid}_{nt}+q_{nt}
         -\sum_{s\in\mathcal{S}_n}P_{st}
\end{equation}
The problem maximizes welfare:
\begin{equation}
\begin{aligned}
  \max\ \sum_t\Big[\,&\tau\sum_{k,r,b}v_{kb}\,d_{krtb}
     -\tau\sum_{\mathcal{A}}c^{\rm net}_{rs}\,z_{krsmt}\\
     &-\sum_i C_i(g_{it})-\sum_n (VOLL) q_{nt}\Big]
\end{aligned}\label{eq:obj}
\end{equation}
We set VOLL = \$50,000/MWh. At this value, we check that every compute bid is below VOLL per MWh, so compute can never outbid other customers’ load. The objective function is subject to, for all $t$ (duals in brackets):
\begin{align}
  &\textstyle\sum_{s,m}z_{krsmt}=\sum_b d_{krtb} && [\lambda_{krt}] \label{eq:bal}\\
  &d_{krtb}\le\bar d_{krtb} \label{eq:dbar}\\
  &\textstyle\sum_{k,r,m}(a_{km}/\rho_k)\,z_{krsmt}\le y_{st} && [\mu^{\rm on}_{st}] \label{eq:on}\\
  &y_{st}\le G_s && [\mu^{\rm G}_{st}] \label{eq:gpu}\\
  &P_{st}\le\bar P_s && [\mu^{\rm P}_{st}] \label{eq:pcap}\\
  &\textstyle\sum_n p_{nt}=0 && [\nu^{\rm ref}_t] \label{eq:grid}\\
  &-F_l\le\textstyle\sum_n H_{ln}\,p_{nt}\le F_l && [\kappa^-_{lt},\kappa^+_{lt}] \label{eq:flow}\\
  &\underline{g}_i\le g_{it}\le\bar g_i \label{eq:gen}\\ 
  & 0 \leq q_{nt} \leq D_{nt}^{\text{grid}}  \label{eq:shed}
\end{align}
The site power is the facility power of busy GPUs plus the idle power $\pi^0$ of all other powered-on GPUs:
\begin{equation}
\begin{aligned}
  P_{st}=\mathrm{PUE}_s\Big[&\textstyle\sum_{k,r,m}e_{km}\,z_{krsmt}\\
   &+\pi^0\big(y_{st}-\textstyle\sum_{k,r,m}a_{km}\,z_{krsmt}\big)\Big]
\end{aligned}\label{eq:pow}
\end{equation}
Power Usage Effectiveness (PUE) is chosen to be a constant multiplier. The constraints are the following:
\begin{itemize}
  \item \eqref{eq:bal}--\eqref{eq:dbar} balance served and accepted requests.
  \item \eqref{eq:on} keeps busy GPUs plus headroom within the powered-on GPUs, while \eqref{eq:gpu} limits to the installed $G_s$.
  \item \eqref{eq:pcap} is the site power limit.
  \item \eqref{eq:grid}--\eqref{eq:shed} are standard DC-OPF constraints in the Power Transfer Distribution Factor (PTDF) formulation. 
\end{itemize}
The rest of the load is shed according to the VOLL. The term $c^{\rm net}_{rs}$ is the tie-breaking cost for routing. 


\subsection{Price of compute}

The dual $\lambda_{krt}$ (\$ per request) of \eqref{eq:bal} is the marginal price of a class-$k$ request in region $r$. The LMP at bus $n$ is
\begin{equation}
\nu_{nt} = \nu^{\mathrm{ref}}_t - \sum_l H_{ln}\left(\kappa^+_{lt} - \kappa^-_{lt}\right)
\end{equation}


Consider site $s$ serving this request in mode $m$, with $n$ as the site's bus. Dropping the indices $k$, $s$, $m$, and $t$, the stationarity conditions for $z$ and $y$ give
\begin{equation}
\lambda = c^{\mathrm{net}} + \frac{1}{\tau}\left[\hat{e}\left(\nu_n + \mu^P\right) + \hat{a}\,\mu^G\right]
\label{eq:compute_price}
\end{equation}
where $\hat{a} = a/\rho$ is the GPU time reserved for the request and $\hat{e} = \mathrm{PUE}\left[e + (\rho^{-1}-1)\pi^0 a\right]$ is its facility energy, including the idle power of its headroom. Thus, a request pays for energy at the local LMP, for the site power limit when it binds, and for GPU time. Sites that do not serve the request have a weakly higher cost, as with generators out of merit. Table~\ref{tab:price_components} separates this price into components that identify the scarce resource.

\begin{table}[t]
\centering
\caption{Components of the compute price, before dividing by $\tau$.}
\label{tab:price_components}
\begin{tabular}{ll}
\hline
Component & Value \\
\hline
Grid energy & $\mathrm{PUE}\,e\,\nu^{\mathrm{ref}}$ \\
Grid congestion & $\mathrm{PUE}\,e\,(\nu_n - \nu^{\mathrm{ref}})$ \\
Site power & $\mathrm{PUE}\,e\,\mu^P$ \\
GPU scarcity & $a\,\mu^G$ \\
SLO headroom & $(\rho^{-1}-1)\,a\left[\mu^G + \mathrm{PUE}\,\pi^0(\nu_n + \mu^P)\right]$ \\
\hline
\end{tabular}
\end{table}

Since $a$ and $e$ are sums of input and output parts, the price splits the same way. For $j \in \{\mathrm{in}, \mathrm{out}\}$, the price per token is
\begin{equation}
\lambda^j = \frac{1}{\tau n^j}\left[\hat{e}^j\left(\nu_n + \mu^P\right) + \hat{a}^j\mu^G\right]
\label{eq:token_price}
\end{equation}
where $\hat{a}^j$ and $\hat{e}^j$ use $a^j$ and $e^j$ in place of $a$ and $e$. The ratio of output to input prices therefore approximately follows $e^{\mathrm{out}}/n^{\mathrm{out}} : e^{\mathrm{in}}/n^{\mathrm{in}}$ when energy or site power is scarce, and $\theta^{\mathrm{pre}}/\theta_m$ when GPUs are scarce.

\section{Experiments}

\subsection{Setup}


We construct a toy example of the co-optimization problem for demonstration purposes.\footnote{The code is available at \url{https://github.com/starship204/ComputePricing}} 



\begin{figure}[t]
  \centering
  \includegraphics[width=0.45\textwidth]{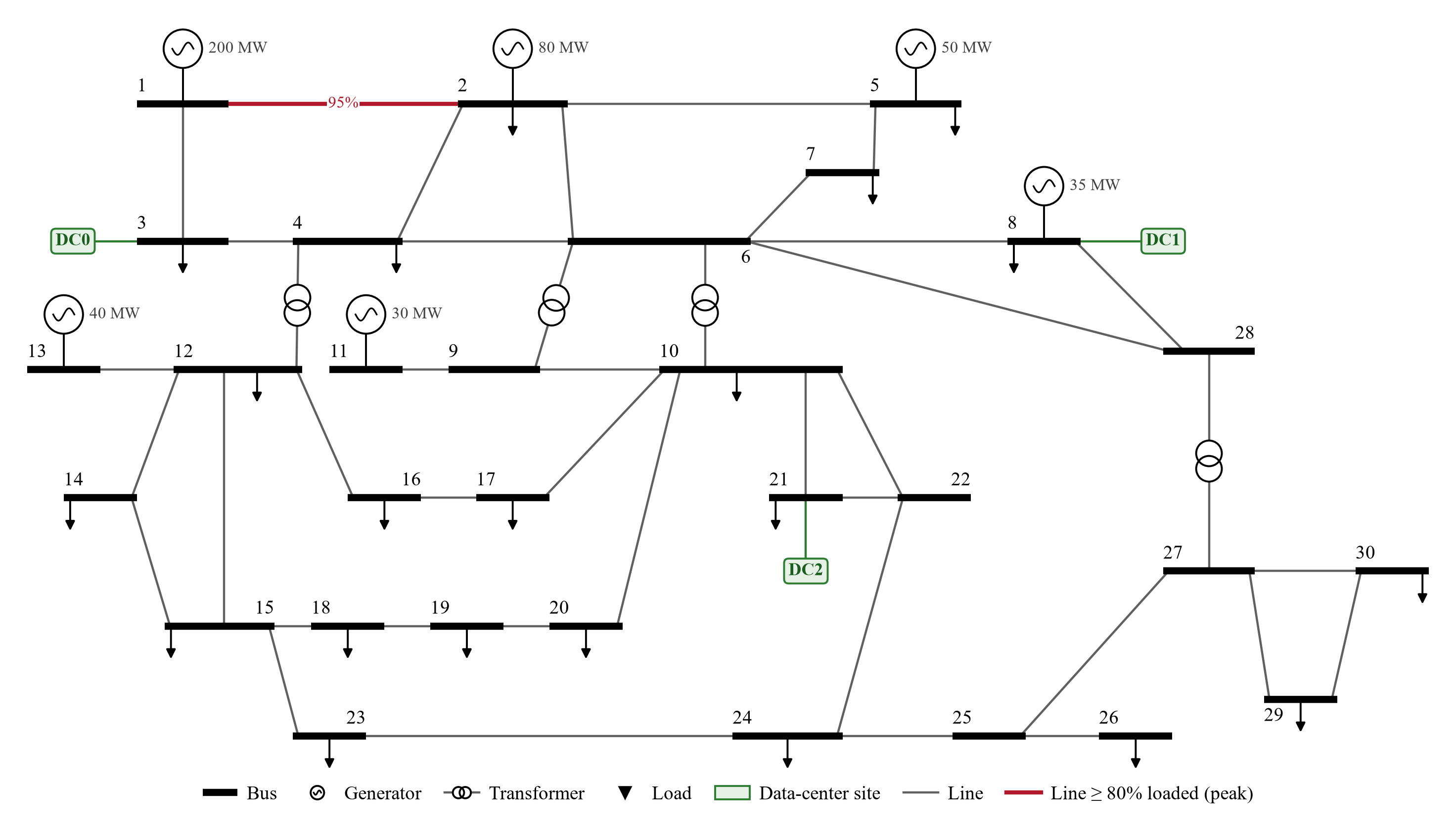}
  \caption{IEEE 30-bus system with the three data-center sites. Red lines are
  loaded at or above 80\% at the peak hour without data centers.}
  \label{fig:grid}
\end{figure}

\textit{Grid.} We use the IEEE 30-bus system from ~\cite{pglib, babaeinejadsarookolaee2019power} in PTDF form. We place three data center sites, DC0, DC1 and DC2, connected at buses 3, 8 and
21 respectively (Fig.~\ref{fig:grid}). Each has a 25\,MW rating, and their PUEs are 1.2, 1.2 and 1.35. Region $r$ is local to site $r$. For the network round-trip time (RTT), we assume 10\,ms to the local site, 60\,ms to a neighboring site, and 120\,ms to the furthest site. The time offsets are not assumed to represent distances on the 30-bus network.



\textit{Compute Load.} The primary link between power and compute comes from ML.ENERGY ~\cite{mlenergy}. We choose Llama-3.1-70B~\cite{llama3} on H100 GPUs with 4-way tensor parallelism. We process tables of $(B_m,\theta_m,e_m,\ell_m)$, quantifying the relationship between power and decode of output tokens, latency, and batch size, and apply a factor of 1.3 for the CPU and additional hardware overhead. We model prefill throughput at 50\% model-FLOPs utilization of the dense BF16 peak, while energy also follows from ~\cite{mlenergy}. We assume that an idle GPU draws 0.12\,kW, and sizing for each site follows $G_s=\lfloor\bar P_s/(\mathrm{PUE}_s P^{\rm GPU})\rfloor$ GPUs, where $P^{\rm GPU}$ is the maximum GPU power draw.


\textit{Demand.} Request rates and lengths come from the Azure 2024 LLM
inference traces~\cite{dynamollm}: conversation for conversation and offline requests, and code for agentic requests. We take the median weekday at 5-minute resolution and average it to hours. Offline demand is flat over the day. The class mix is 50/30/20\%. Regions take 50/30/20\% of demand and are shifted by 0, 1, and 2 hours. Demand is scaled so that, at the peak, the GPUs needed, including headroom, fill 90\% of the fleet. The latency targets enter the clearing as $z_{krsmt}=0$ for items outside \eqref{eq:adm}. 



\textit{Events.} We study three representative events. At DC1 during 14:00–20:00 UTC: ``GPU failure" represents 60\% of GPUs offline, ``power limit" represents reduced site power from 25 to 10 MW, and ``grid event" limits the lines into bus 8 to 30\% of their rating.



\begin{table}[t]
\centering
\caption{Request classes. Bids use $\varepsilon=1$ and $N_B=10$ segments.}
\label{tab:classes}
\setlength{\tabcolsep}{3.5pt}
\begin{tabular}{lccc}
\toprule
                               & Conversation & Agentic & Offline \\
\midrule
Trace for lengths              & conv.       & code    & conv.   \\
$n^{\rm in}_k\,/\,n^{\rm out}_k$ (ratio) & 16 & 114 & 16 \\
Share of requests              & 50\%        & 30\%    & 20\%    \\
ITL target $L_k$ (ms)          & 30          & 50      & --      \\
RTT budget $R_k$ (ms)          & 50          & 100     & --      \\
Wait target $W_k$ (s)          & 1           & 10      & --      \\
Utilization $\rho_k$           & 0.85   & 0.98 & 0.95    \\
Bid range (\$/M tokens)        & 0.40--3.00  & 0.40--3.00 & 0.20--0.45 \\
\bottomrule
\end{tabular}
\end{table}


\begin{table}[t]
\caption{Results at the peak hour. Prices are for region 1 in \$/M-tokens; served shares are fleet-wide.}
\label{tab:quantitative_results}
\centering
\scriptsize
\setlength{\tabcolsep}{3pt}
\renewcommand{\arraystretch}{1.15}
\begin{tabular}{lrrrr}
\toprule
\textbf{Event} & \textbf{Base} & \textbf{GPU failure} &
\textbf{power limit} & \textbf{grid event} \\
\midrule
LMP at DC1 (\$/MWh) & 56.6 & 55.3 & 55.9 & 2,413.3 \\
DC1 power (MW) & 21.8 & 7.66 & 10.0 & 12.1 \\
\midrule
Conversation price & 0.007 & 0.400 & 0.316 & 0.270 \\
Agentic price & 0.005 & 0.235 & 0.229 & 0.197 \\
Offline price & 0.006 & 0.262 & 0.255 & 0.219 \\
\midrule
Conversation served (\%) & 100 & 90.6 & 96.9 & 100 \\
Agentic served (\%) & 100 & 100 & 100 & 100 \\
Offline served (\%) & 100 & 72.3 & 76.3 & 86.2 \\
\midrule
Welfare loss (k\$/day) & 0 & 75.5 & 34.5 & 5.03 \\
\bottomrule
\end{tabular}
\end{table}

\subsection{Results}

\begin{figure}[t]
  \centering
  \includegraphics[width=0.48\textwidth]{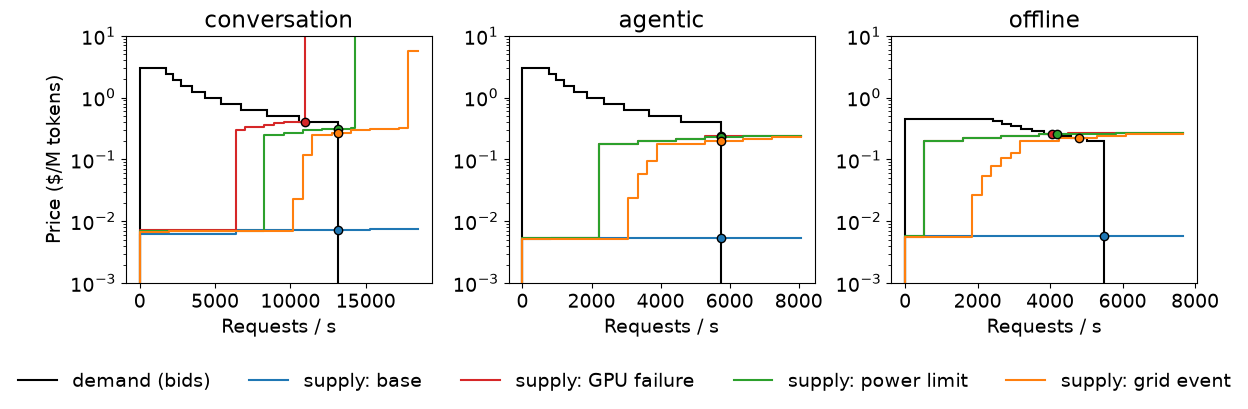}
  \caption{Demand bids (black) and residual supply (colored) for region 1 at the peak hour, per event. Intersections mark the cleared quantity and price.}
  \label{fig:clearing}
\end{figure}

\textit{Clearing.} Fig.~\ref{fig:clearing} shows that supply is flat at the marginal cost of energy for the base case. The events create supply scarcity. The clearing moves onto the bids: the price is set by the willingness to pay of the marginal segment, and the lowest-value segments are curtailed. Table~IV summarizes the peak hour results and their clearing.

\begin{figure}[t]
  \centering
  \includegraphics[width=0.48\textwidth]{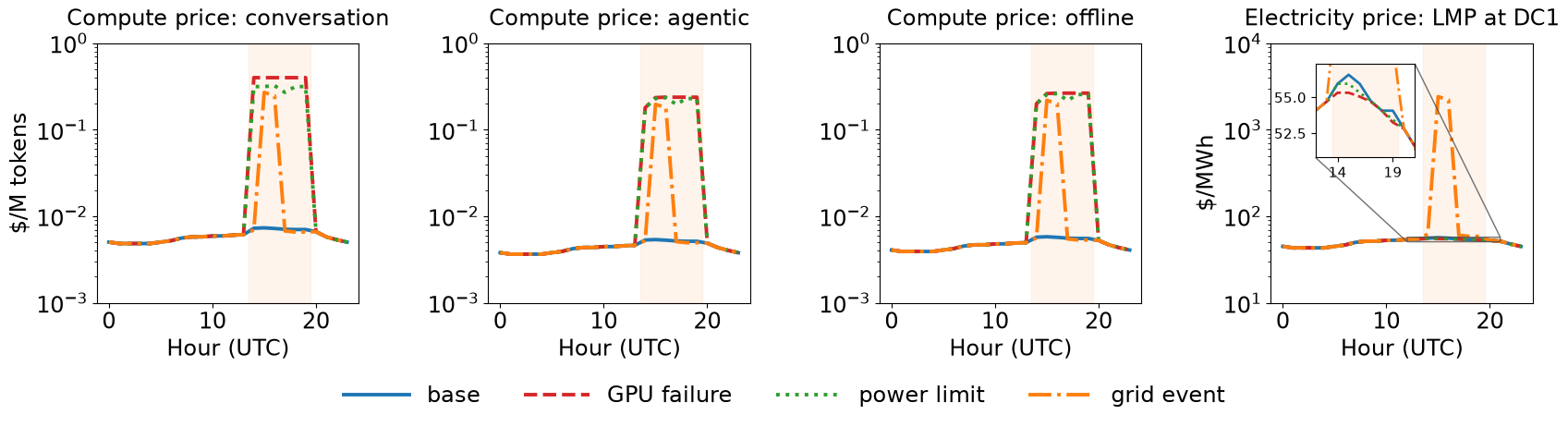}
  \caption{Compute prices in region 1 and the electricity LMP at the DC1 bus over the day. The shaded area is the event window.}
  \label{fig:priceday}
\end{figure}

\textit{Prices over the day.} Fig.~\ref{fig:priceday} shows that, outside the event window, all prices stay at the marginal cost of energy. Within the window, prices rise only in the hours when DC0 and DC2 can no longer absorb the work moved from DC1. The surge therefore lasts longest for the GPU failure, which removes the most capacity, and shortest for the grid event. The three compute prices follow nearly the same path because offline work absorbs most of the curtailment: its bid sets the value of the scarce resource, and each class pays the value in proportion to its use.

\begin{figure}[t]
  \centering
  \includegraphics[width=0.45\textwidth]{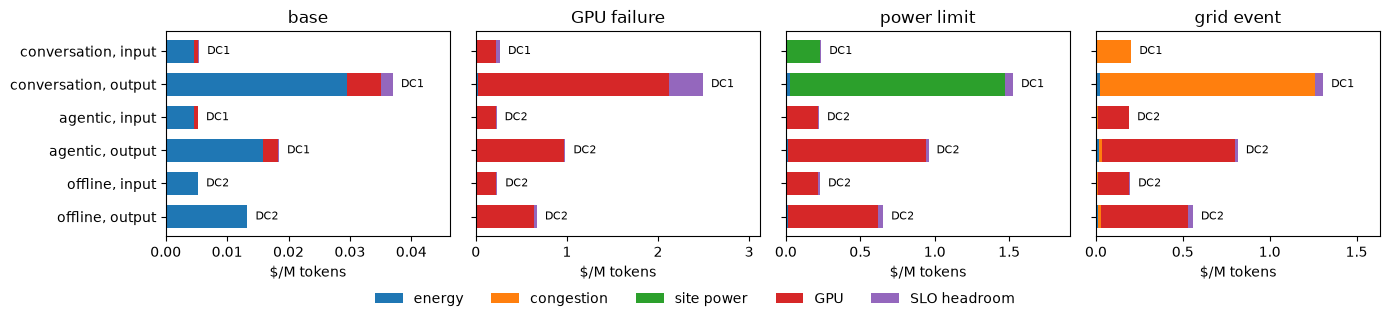}
  \caption{Components of the input- and output-token prices in region 1 at the
  peak hour, per event. Labels give the site serving most of each class.}
  \label{fig:decomp}
\end{figure}

\textit{Decomposition.} Fig.~\ref{fig:decomp} shows which constraints bind. In the base case, every price is almost entirely energy. During the events, conversation from region 1 is still served only at DC1. Its 50 ms round-trip budget is met only by the local site, since DC0 and DC2 are 60 ms away. Its price therefore reveals the constraint that binds at DC1: GPUs in the GPU failure, site power in the power limit, and congestion in the grid event. Agentic and offline requests from region 1 move to DC0 and DC2 instead. The moved work fills the GPUs at DC0 and DC2, and once those sites are full, the lowest offline bids are curtailed and set the value of a GPU-hour there. GPU scarcity therefore dominates their prices in every event, including the grid event, where DC1 is limited by grid congestion.

\section{Discussion \& Conclusion}




The proposed approach provides a demonstrative example of how the compute price reflects the binding constraints of both the grid and the compute infrastructure. Nevertheless, the concept poses several challenges. First, the compute price is marginal: it prices the short-run use of infrastructure and does not recover capital costs, which dominate GPU and data center economics. Capital recovery must therefore be addressed separately. Second, the approach requires transparency and collaboration between the grid operator and the compute operator. The two sides could potentially clear separately by exchanging prices and quantities rather than full models.



While we acknowledge these practical limitations, we hope to introduce a paradigm in which compute and energy are coordinated rather than disaggregated systems. By grounding the compute price in economic merit, the approach introduces the potential of a transparent way to allocate scarce infrastructure and to turn data center flexibility into a signal that users can see and act on. Moreover, without the network, the same model prices compute for off-grid or behind-the-meter sites.

Many extensions remain. On the grid side, these include AC-OPF formulations, larger fleets and realistic transmission networks, and integration with wholesale market operations at scale. Moreover, each hour is cleared independently, so the price signals when deferral is valuable, but deferral itself is not modeled. On the compute side, they include more rigorous serving constraints, finer compute--power coupling that captures latency distributions, and commitment decisions with appropriate pricing. Finally, dedicated clearing algorithms, possibly in a decentralized way, would be needed to solve the problem at scale.

\section*{Acknowledgment}

The authors disclose the use of Anthropic Claude to conduct this work. The usage included assistance in writing, literature review, methodology refinement, and code generation and editing. Conceptual ideas, experimental design, implementations, analyses, and conclusions are the authors’ own. The authors take responsibility for all mistakes.






\bibliographystyle{IEEEtran}
\bibliography{citations}

@inproceedings{dynamollm,
  title={Dynamollm: Designing llm inference clusters for performance and energy efficiency},
  author={Stojkovic, Jovan and Zhang, Chaojie and Goiri, {\'I}{\~n}igo and Torrellas, Josep and Choukse, Esha},
  booktitle={2025 IEEE International Symposium on High Performance Computer Architecture (HPCA)},
  pages={1348--1362},
  year={2025},
  organization={IEEE}
}

@article{babaeinejadsarookolaee2019power,
  title={The power grid library for benchmarking ac optimal power flow algorithms},
  author={Babaeinejadsarookolaee, Sogol and Birchfield, Adam and Christie, Richard D and Coffrin, Carleton and DeMarco, Christopher and Diao, Ruisheng and Ferris, Michael and Fliscounakis, Stephane and Greene, Scott and Huang, Renke and others},
  journal={arXiv preprint arXiv:1908.02788},
  year={2019}
}

@misc{pglib,
  author       = {{IEEE PES Task Force on Benchmarks for Validation of Emerging Power System Algorithms}},
  title        = {{Power Grid Lib -- Optimal Power Flow (PGLib-OPF)}},
  year         = {2024},
  howpublished = {\url{https://github.com/power-grid-lib/pglib-opf}},
  note         = {Accessed: November 5, 2026}
}

@article{llama3,
  title={The llama 3 herd of models},
  author={Grattafiori, Aaron and Dubey, Abhimanyu and Jauhri, Abhinav and Pandey, Abhinav and Kadian, Abhishek and Al-Dahle, Ahmad and Letman, Aiesha and Mathur, Akhil and Schelten, Alan and Vaughan, Alex and others},
  journal={arXiv preprint arXiv:2407.21783},
  year={2024}
}

@article{mlenergy,
  title={The ML. ENERGY benchmark: Toward automated inference energy measurement and optimization},
  author={Chung, Jae-Won and Ma, Jeff J and Wu, Ruofan and Liu, Jiachen and Kweon, Oh Jun and Xia, Yuxuan and Wu, Zhiyu and Chowdhury, Mosharaf},
  journal={Advances in Neural Information Processing Systems},
  volume={38},
  year={2026}
}

@article{liu2026peer,
  title={Peer-to-Peer Cloud Service Market for Data Centers Oriented to Computation-Electricity Coordination},
  author={Liu, Yugui and Ding, Yibo and Li, Xudong and Qu, Jing and Zhang, Wenyi and Qian, Tong and Xiao, Wuyou and Hu, Zhengyang and Ruan, Jiaqi and Xu, Zhao},
  journal={arXiv preprint arXiv:2606.04981},
  year={2026}
}

@article{zhang2022remunerating,
  title={Remunerating space--time, load-shifting flexibility from data centers in electricity markets},
  author={Zhang, Weiqi and Zavala, Victor M},
  journal={Applied Energy},
  volume={326},
  pages={119930},
  year={2022},
  publisher={Elsevier}
}

@article{ding2026towards,
  title={Towards cost-optimal joint electricity-computation management: A novel predict-then-optimize framework},
  author={Ding, Yibo and Li, Xudong and Zhao, Yuhong and Shi, Wenzhuo and Lyu, Cheng and Ruan, Jiaqi and Xu, Zhao},
  journal={Applied Energy},
  volume={412},
  pages={127734},
  year={2026},
  publisher={Elsevier}
}

@article{sanabria2026shadow,
  title={The Shadow Price of Intelligence: Quality Degradation in LLM Inference as a Supply Chain Problem},
  author={Sanabria, Elioth},
  journal={arXiv preprint arXiv:2608.23986},
  year={2026}
}

@article{du2026spatial,
  title={Spatial LLM Workload Shifting Needs Foresight: Model Commitment for AI Data Center Operation under Power Grid Constraints},
  author={Du, Bojun and Jia, Hongyang and Li, Tonghui and Hou, Qingchun and Wang, Ze and Du, Ershun and Zhang, Ning},
  journal={arXiv preprint arXiv:2609.09787},
  year={2026}
}

@incollection{liu2026locational,
  title={Locational Pricing for Generative-AI Services via Token-Flow Market Clearing},
  author={Liu, Shaohui},
  booktitle={Proceedings of the 2026 ACM Sustainability Week},
  pages={258--265},
  year={2026}
}

@article{billimoria2026bits,
  title={Bits-to-Watts: Connecting Markets and Prices for Compute and Power},
  author={Billimoria, Farhad and Byers, Conleigh},
  year={2026}
}

@article{liu2026watts,
  title={Watts vs. Bytes: Turning Data Centers into Grid Assets via Storage Compute Co-Optimization},
  author={Liu, Shaohui and Shin, Sungho and Deka, Deepjyoti},
  journal={arXiv preprint arXiv:2605.16190},
  year={2026}
}

@article{gu2025competitive,
  title={Competitive Equilibrium for Electricity Markets with Spatially Flexible Loads},
  author={Gu, Nan and Qin, Junjie},
  journal={arXiv preprint arXiv:2510.26036},
  year={2025}
}

@inproceedings{lin2021evaluating,
  title={Evaluating coupling models for cloud datacenters and power grids},
  author={Lin, Liuzixuan and Zavala, Victor M and Chien, Andrew A},
  booktitle={Proceedings of the Twelfth ACM International Conference on Future Energy Systems},
  pages={171--184},
  year={2021}
}

@article{zhang2020flexibility,
  title={Flexibility from networks of data centers: A market clearing formulation with virtual links},
  author={Zhang, Weiqi and Roald, Line A and Chien, Andrew A and Birge, John R and Zavala, Victor M},
  journal={Electric Power Systems Research},
  volume={189},
  pages={106723},
  year={2020},
  publisher={Elsevier}
}

@article{lindberg2022using,
  title={Using geographic load shifting to reduce carbon emissions},
  author={Lindberg, Julia and Lesieutre, Bernard C and Roald, Line A},
  journal={Electric Power Systems Research},
  volume={212},
  pages={108586},
  year={2022},
  publisher={Elsevier}
}

@inproceedings{lindberg2021guide,
  title={A guide to reducing carbon emissions through data center geographical load shifting},
  author={Lindberg, Julia and Abdennadher, Yasmine and Chen, Jiaqi and Lesieutre, Bernard C and Roald, Line},
  booktitle={Proceedings of the Twelfth ACM International Conference on Future Energy Systems},
  pages={430--436},
  year={2021}
}

@article{liu2013data,
  title={Data center demand response: Avoiding the coincident peak via workload shifting and local generation},
  author={Liu, Zhenhua and Wierman, Adam and Chen, Yuan and Razon, Benjamin and Chen, Niangjun},
  journal={Performance Evaluation},
  volume={70},
  number={10},
  pages={770--791},
  year={2013},
  publisher={Elsevier}
}

@article{liu2011greening,
  title={Greening geographical load balancing},
  author={Liu, Zhenhua and Lin, Minghong and Wierman, Adam and Low, Steven H and Andrew, Lachlan LH},
  journal={ACM SIGMETRICS Performance Evaluation Review},
  volume={39},
  number={1},
  pages={193--204},
  year={2011},
  publisher={ACM, New York, NY, USA}
}

@inproceedings{rao2010minimizing,
  title={Minimizing electricity cost: Optimization of distributed internet data centers in a multi-electricity-market environment},
  author={Rao, Lei and Liu, Xue and Xie, Le and Liu, Wenyu},
  booktitle={2010 Proceedings IEEE INFOCOM},
  pages={1--9},
  year={2010},
  organization={IEEE}
}

@article{fatima2026unlocking,
  title={Unlocking AI Data Center Interconnection Capacity Through Coordinated Grid and Data Center Flexibility},
  author={Fatima, Rida and Li, Xingpeng},
  journal={arXiv preprint arXiv:2610.05502},
  year={2026}
}

@article{radovanovic2022carbon,
  title={Carbon-aware computing for datacenters},
  author={Radovanovi{\'c}, Ana and Koningstein, Ross and Schneider, Ian and Chen, Bokan and Duarte, Alexandre and Roy, Binz and Xiao, Diyue and Haridasan, Maya and Hung, Patrick and Care, Nick and others},
  journal={IEEE Transactions on Power Systems},
  volume={38},
  number={2},
  pages={1270--1280},
  year={2022},
  publisher={IEEE}
}

@inproceedings{sukprasert2024limitations,
  title={On the limitations of carbon-aware temporal and spatial workload shifting in the cloud},
  author={Sukprasert, Thanathorn and Souza, Abel and Bashir, Noman and Irwin, David and Shenoy, Prashant},
  booktitle={Proceedings of the Nineteenth European Conference on Computer Systems},
  pages={924--941},
  year={2024}
}

@incollection{chen2026defer,
  title={To Defer or To Shift? The Role of AI Data Center Flexibility on Grid Interconnection},
  author={Chen, Yize and Zheng, Xiaogui},
  booktitle={Proceedings of the 2026 ACM Sustainability Week},
  pages={322--327},
  year={2026}
}

@article{colangelo2026ai,
  title={AI data centres as grid-interactive assets},
  author={Colangelo, Philip and Coskun, Ayse K and Megrue, Jack and Roberts, Ciaran and Sengupta, Shayan and Sivaram, Varun and Tiao, Ethan and Vijaykar, Aroon and Williams, Chris and Wilson, Daniel C and others},
  journal={Nature Energy},
  volume={11},
  number={2},
  pages={254--261},
  year={2026},
  publisher={Nature Publishing Group UK London}
}

@article{bogmans2026power,
  title={Power hungry: How ai will drive energy demand},
  author={Bogmans, Christian and Ganpurev, Ganchimeg and Gomez-Gonzalez, Patricia and Melina, Giovanni and Pescatori, Andrea and Thube, Sneha},
  journal={Energy Economics},
  pages={109278},
  year={2026},
  publisher={Elsevier}
}

@article{cheng2026ai,
  title={AI for the grid and AI on the grid: Implications, challenges and opportunities},
  author={Cheng, Bo and Botterud, Audun and Levin, Todd and Nadarajah, Selvaprabu and Zhao, Dongwei and Kwon, Jonghwan},
  journal={The Electricity Journal},
  volume={39},
  number={2},
  pages={107547},
  year={2026},
  publisher={Elsevier}
}

@article{chen2025electricity,
  title={Electricity demand and grid impacts of AI data centers: Challenges and prospects},
  author={Chen, Xin and Wang, Xiaoyang and Colacelli, Ana and Lee, Matt and Xie, Le},
  journal={arXiv preprint arXiv:2509.07218},
  year={2025}
}

@article{shehabi20242024,
  title={2024 united states data center energy usage report},
  author={Shehabi, Arman and Newkirk, Alex and Smith, Sarah J and Hubbard, Alex and Lei, Nuoa and Siddik, Md Abu Bakar and Holecek, Billie and Koomey, Jonathan and Masanet, Eric and Sartor, Dale},
  year={2024}
}


\end{document}